\documentclass[superscriptaddress,twocolumn,showpacs,preprintnumbers,amsmath,amssymb]{revtex4}

\def\beq{\begin{equation}}
\def\enq{\end{equation}}
\def\bea{\begin{eqnarray}}
\def\ena{\end{eqnarray}}
\def\<{<\!\!}
\def\>{\!\!>}
\def\siml{\lower.5ex\hbox{$\; \buildrel < \over \sim \;$}}
\def\simg{\lower.5ex\hbox{$\; \buildrel > \over \sim \;$}}

\def\Mesz{M\'esz\'aros~}
\def\prl{Phys.\ Rev.\ Lett.~}
\def\prd{Phys.\ Rev.\ D~}
\def\plb{Phys.\ Lett.\ B~}

\def\apj{Astrophys.\ J.~}
\def\aa{Astron.\ Astrophys.~}
\def\mnras{Mon.\ Not.\ R.\ Astron.\ Soc.~}
\def\bitm{\bibitem}
\def\mpl{m_{Pl}}
\def\tpl{t_{Pl}}
\def\Epl{E_{Pl}}
\def\lpl{\ell_{Pl}}
\def\s{\hbox{s}}
\def\g{\hbox{g}}
\def\GeV{\hbox{GeV}}
\def\Mbh{M_{BH}}
\def\Mh{M_{Hor}}
\def\tbh{t_{BH}}
\def\Tbh{T_{BH}}
\def\T300{T_{ew,300}}
\def\gast{g_\ast}
\def\teq{t_{eq}}
\def\Omh{\Omega_{M,0}h^2}

\input{epsf}
\begin{document}
\title{Reheating, Dark Matter and Baryon Asymmetry: \\ 
a Triple Coincidence in Inflationary Models}

\author{Stephon Alexander}
\author{Peter  M\'esz\'aros}

\affiliation{Department of Physics and Department of Astronomy \& Astrophysics, \\
Pennsylvania State University, University Park, PA 16802}

\begin{abstract}
A scenario in which primeval black holes (PBHs) form at the end of an
extended inflationary period is capable of producing, via Hawking radiation, 
the observed entropy, as well as the observed dark matter density in the 
form of Planck mass relics. The observed net baryon asymmetry is produced by 
sphaleron processes in the domain wall surrounding the PBHs as they evaporate 
around the electroweak transition epoch. The conditions required to satisfy 
these three observables determines the PBH formation epoch, which can be 
associated with the end of inflation, at $t\sim 10^{-32}\s$.
\end{abstract}


\date{\today}
\maketitle

\section{Introduction}

The possibility of primeval black hole (PBH) formation 
in the early Universe has been known for a long time \cite{zeld67,haw71}. The
realization that quantum effects lead to their evaporation \cite{haw74}
led to investigations of possible constraints on the PBH mass spectrum 
from their consequences for various astrophysical backgrounds
\cite{carr75,carr76} and their possible dynamical role as dark matter \cite{mesz75}. 
The interplay between PBHs as a source of CDM affecting cosmological dynamics, 
and their evaporation as a source of entropy and particles \cite{page76}
affecting nucleosynthesis and CMB observations leads to constraints on their 
mass spectrum over a wide mass range \cite{sato78,nov79}. Since black 
hole evaporation  manifestly violates T-invariance, CP must be violated as
well so that baryon number is not conserved \cite{haw76}, which led to the 
suggestion, in the context of GUTs, that the baryon asymetry of the Universe 
could be thereby explained \cite{CP-GUT79,tur79}. In the context of inflationary
models, PBH formation can be triggered by large amplitude inhomogeneities caused
by bubble nucleation over scales comparable to the horizon, which can collapse
into black holes at the end of an extended inflationary period \cite{bar91-92}. 

The evaporation of PBHs with mass less than $\sim 10^{15}$ g, on timescales 
less than a Hubble time \cite{haw74}, can lead either to complete evaporation, 
or may stop at a Planck scale mass $\mpl\sim 10^{-5}$ g \cite{relics}.
Much work has been done since then on the possibility of PBH formation at
phase transitions, on the dynamics of their collapse and on their possible role 
in large scale structure formation, e.g. \cite{pbh99-06,carr05} and references 
therein.

In what follows, we concentrate on the potential of PBHs formed at the end of 
extended inflation in providing a mechanism for the production of the current
observed entropy per baryon, the inferred dark matter density, and as a possible
source for the baryon asymmetry in the Universe.  We emphasize a remarkable
triple coincidence for the conditions required to produce these three
quantities, pointing towards a well defined epoch for PBH formation around 
$t\sim 10^{-32}\s$, which can be identified with the end of inflation.

In this model, the absolute entropy  of the universe is given by the entropy
of a gas of standard model particles at the initial temperature $\Tbh\sim 300
\GeV$, produced by PBHs created at $t\sim 10^{-32}\s$ which evaporate at 
$\tbh\sim 10^{-12}\s$, identified with the reheating time. This same PBH 
evaporation leads also to a dark matter component, assumed to consist of 
approximately Planck mass remnants with a mass density approximately equal 
to the PBH density times the Planck mass.  The net baryon density and asymmetry 
is also related to the PBH evaporation, through sphaleron processes in a domain 
wall structure surrounding the PBHs.  The presently observed entropy, dark 
matter density and net baryon to entropy ratio are obtained for a unique value 
of the reheating time $\tbh\sim 10^{-12}\s$, which coincides with the
electroweak timescale, and which defines a unique time for the end of inflation at $t_{end}
\sim 10^{-32}\s$.

\section{PBHs, Reheating and Entropy}
\label{sec:entropy}

At the Planck time  $\tpl = (\hbar G/c^5)^{1/2}\sim 10^{-43.3}$ s the Planck 
mass $\mpl= (\hbar/\tpl c^2)\sim (\hbar c^3/G)\sim 10^{-4.7}\g$, corresponding
to the Planck energy $\Epl\sim 10^{19.05}\GeV$, is within a particle horizon 
whose size is the Planck length $\lpl\sim c\tpl \sim 10^{-32.83}$ cm.
Any PBHs formed before or during inflation would have had their energy density 
diluted by the exponential expansion of the scale factor to a negligible 
value, so that PBH formation is of interest mainly after the end of inflation, 
at $t\simg t_{end}$.  

The difficulties of the original inflation model are 
resolved most simply in models of extended or hyperextended inflation
\cite{extinfl}, which is generally taken to end around an epoch $t_{end} 
\sim 10^{-32 \pm 6}$ s, at which time the energy scale has dropped to 
$E\sim \rho^{1/4}\sim 10^{13 \pm 3}$ GeV. At this time the Universe is cold,
due to adiabatic cooling during the expansion of the scale factor by sixty or 
more e-foldings, so the pressure is essentially zero, and the equation of 
state is correspondingly soft. Primordial energy density fluctuations coming
into the horizon at $t\sim t_{end}$ may be of the canonical inflationary 
(Harrison-Zeldovich) type, with relative amplitudes $\delta_{end}\equiv 
(\delta\rho/\rho)_{t_{end}}\sim 10^{-4}$, and/or may be large amplitude 
fluctuations caused by chaotic conditions associated with bubble nucleation
at $t_{end}$, where one may expect $\delta_{end}\sim 1$, e.g. 
\cite{bar91-92}.  The latter fluctuations can cause PBHs to form almost 
immediately at $t_1\sim t_{end}$, with a mass $\Mbh$ which is a fraction 
$\eta\siml 1$ of the mass in the horizon $M_{\rm hor}\sim \mpl(t/\tpl)$ at 
that time,
\beq
\Mbh (t_1) \simeq \eta \mpl (t_1/\tpl) \simeq 10^{6.6} \eta t_{1,-32}~\g
\label{eq:Mbh}
\enq
or $\Mbh\simeq 10^{30.3} \eta t_{1,-32}~\GeV$, where $t_{1,-32} =
(t_1/10^{-32}\s)$.
The temperature associated with a black hole of mass $\Mbh$ is 
%$k\Tbh=\hbar c^3/8\pi GM)
\bea
\Tbh &=&({\mpl^2}/{8\pi\Mbh}) \nonumber\\
     &=&10^{6.4}M_{6.6}^3\GeV= 10^{6.4}\eta^{-1}t_{1,-32}\GeV.
\label{eq:Tbh}
\ena
In the standard treatment, these PBHs evaporate on a timescale 
\bea
\tbh (\Mbh) &=& g_\ast^{-1} (\Mbh/\mpl)^3 \tpl  \nonumber\\
            &\simeq& 10^{-11.4} g_{\ast,2}^{-1} (\eta t_{1,-32})^3 ~\s,
\label{eq:tbh}
\ena
where $g_\ast=10^2 g_{\ast,2} \sim 106.75$ is the number of degress of freedom 
in the early universe for the standard model\cite{kt90}. Most of the 
evaporated energy goes into radiated photons and particles. 

For evaporation times much longer than the epoch of formation, $\tbh \gg t_1$,
the epoch (age of the Universe), at which PBHs of mass $\Mbh$ evaporate is 
$t\sim \tbh \sim 10^{38}\Mbh ~\s$.  Thus even if the perturbations coming into the horizon at the end of inflation 
had only the canonical amplitude $\delta_{end} \sim 10^{-4}$, they grow 
with the scale factor of the Universe $a$ as $\delta\propto a \propto t^{2/3}$ 
(for a matter dominated [MD] Universe, if the equation of state is cold).
%as $\delta\propto a^2 \propto t$ (for a radiation dominated [RD] Universe). 
The collapse time $t_{col}$ at which the fluctuations achieve large amplitude 
$\delta\sim 1$ is much smaller than $\tbh$, and for fluid-like  perturbations
the epoch at which PBHs of mass $\Mbh$ evaporate is again $t\sim \tbh$. 

If $\beta(\Mbh)$ is the fraction of the energy density of the Universe which 
collapses into PBHs of mass $\sim \Mbh$ at the the epoch $t_1$, the radiation 
produced by the PBHs at the evaporation time $\tbh$, after having relaxed with 
the environment, results in a specific entropy per baryon $S=s/n_B$ of 
$S\simeq (1+S_i)\beta(M)(M/\mpl)$ \cite{nov79}, where $S_i$ is the initial 
entropy per baryon before PBH evaporation, assuming $\beta \ll 1$. This can be 
used to set constraints on the fraction $\beta$ of PBHs of mass $M$, and can 
also contribute to producing some or possibly most of the entropy of the 
universe. Generalizing this argument to an inflationary scenario 
\cite{bar91-92}, with $S_{i} \simeq 0$ as expected from adiabatic cooling at 
$t_1\sim t_{end}$, assuming that the PBH mass is a fraction $\eta \Mh$ of the 
mass in the horizon at the collapse time $t_1$, the initial energy density in 
a PBH component is
\beq 
\rho_{BH}(t_1)=(3/32\pi)\beta (\mpl/t_1^3)(t_1/\tpl),
\label{eq:rhobht1}
\enq
while the remaining fraction $(1-\beta)$ goes into relativistic particles or
radiation, $\rho_R(t_1)$. For plausible values of $\eta\siml 1$, 
$10^{-10}\siml \beta\siml 1$, the universe expansion is initially dominated by 
radiation, $a\propto t^{1/2}$, but after a time $t_2=[(1-\beta)^1/\beta^2]t_1$
it becomes PBH dominated \cite{bar91-92}, $a\propto t^{2/3}$. The PBHs 
evaporate at $t\simeq \tbh \gg t_1$, injecting into the universe a radiation 
energy density $\rho_R(\tbh)= (3/32\pi) (1-\beta) \eta^{-6} \gast^2
(\tpl/t_1)^3 (\mpl/t_1^3)$, which can be re-expressed as a function of the 
initial mass of the PBHs which evaporate at $\tbh$,
\beq
\rho_R(\tbh)=(3/32\pi)(1-\beta)(\mpl/\tpl^3)(\mpl/\Mbh)^2~.
\label{eq:rhoradtbh}
\enq
This newly injected radiation component is much larger than the diluted
radiation produced at time $t_1$, and it provides henceforth the dominant energy
form in the universe, which again expands as $a\propto t^{1/2}$ (until
$t_{eq}$). 
A this time the universe acquires an entropy density $s(\tbh)=(2\pi^2/45)\gast 
T(\tbh)^3$, being reheated to a temperature $T(\tbh)$ given by
\bea
T(\tbh) &=& (30/\gast \pi^2)^{1/4}\rho_R(\tbh)^{1/4}  \nonumber\\
  &=&\left(\frac{90\gast [1-\beta]}{32\pi^3}\right)^{1/4}
  \left(\frac{\mpl}{\tpl^3}\right)^{1/4}\left(\frac{\mpl}{\Mbh}\right)^{3/2}
\nonumber\\
  &\simeq& 250 g_{\ast,2}^{1/4}(1-\beta)^{1/4}M_{6.6}^{-3/2}~\GeV
\label{eq:Ttbh}
\ena
Taking the standard value  for the matter-radiation equilibrium epoch 
%$\teq= 4.2\times 10^{10}(\Omh)^{-2}\s$, 
$a_{eq}=4.3\times 10^{-5}(\Omh)^{-1}$, the PBH evaporation epoch corresponds 
to $a_{BH}=4.1\times 10^{-16}M_{6.6}^{3/2}$, and from the entropy scaling 
$T^3 a^3 \gast =$ constant with $g_{\ast,BH}\simeq 106$ and $g_{\ast,0}=3.9$ 
one obtains a present day radiation temperature $T_0 \simeq 3.0\times 
10^{-4} (1-\beta)^{1/4} \hbox{eV}$, close to the observed value of 
$2.5\times 10^{-4}\hbox{eV}$ (See Fig. 1). 
Notice that if one were only trying to explain the current entropy or the 
current radiation temperature, one could in principle also satisfy this 
with, e.g. earlier evaporation times or higher $\Tbh$ values.  However, if 
in addition one demands that the PBH evaporation should also lead to the 
currently observed dark matter density, the evaporation time becomes
determined, as discussed below.

\section{Dark Matter}
\label{sec:dm}

The evaporation timescale (\ref{eq:tbh}) remains approximately the same even 
if mass loss stops after the PBH has shrunk down to a Planck mass $\sim 10^{-5}$
g, since by this time it will have radiated away most of its energy in the 
form of photons and particles. The semi-classical quantum evaporation treatment
breaks down near $\sim \mpl$, requiring taking quantum gravity effects into
account, and several authors have argued that the process leaves behind stable 
relics of approximately a Planck mass \cite{relics}. These would behave as
non-relativistic matter, and in order not to exceed limits on the current dark 
matter density, they imply constraints on the epoch at which they evaporated, 
and therefore also on the epoch at which they formed \cite{bar91-92,carr05}. 
At the epoch $\tbh$ when they evaporate, if each PBH leaves a relic of mass 
$\kappa\mpl$, the relic matter density $\rho_{M}$ is
\beq
\rho_{M}=(3/32\pi)(1-\beta)\gast^2(\mpl/\Mbh)^7 (\kappa\mpl/\tpl^3)
\label{eq:rhoreltbh}
\enq
At this time $\tbh$ the PBH-contributed radiation density (\ref{eq:rhoradtbh})
is dominant, and the ratio of radiation (including relativistic particles) 
density to relic (dark) matter density is
\beq
(\rho_R/\rho_{M})_{\tbh} \simeq (\Mbh /\kappa \mpl) \equiv (\eta
t_1/\kappa\tpl),
\label{eq:massratiotbh}
\enq
which is $\simeq 2\times 10^{11} \kappa^{-1}M_{6.6} = 2\times 10^{11}
\kappa^{-1} \eta t_{1,-32}$. This ratio decreases as $a^{-1}$, and at $\teq$ 
its value is 
\bea
(\rho_R/\rho_{M})_{t_{eq}} &\simeq& 2 M_{6.6}^{5/2}\kappa^{-1}\Omh \nonumber\\
                            &\simeq& 2 t_{1,-32}^{5/2}\eta\kappa^{-1}\Omh ~,
\label{eq:massratioteq}
\ena
close to unity, as needed for dark matter.  The evaporation at $\tbh\sim 
10^{-11.4}~\s$ of PBHs formed at $t\sim 10^{-32}\s$ leads therefore to a 
plausible model for explaining the reheating of the Universe after the end of 
inflation, leading to the right amount of present day entropy, as well as 
providing a source for the present day dark matter density. The latter 
could be in the form of stable Planck mass relics, or possibly stable, weakly
interacting decay products of such relics, with the same total mass.  This is 
achieved if (i) the end of inflation occurs at $t_{end}\sim 10^{-32}\s$,
(ii) PBHs collapse is dominated by fluctuations coming into the horizon at
$t_1\sim t_{end}$, aided by the soft equation after the end of inflation, and 
iii) the fraction of the energy density of the Universe collapsing into PBHs 
at that epoch is $10^{-10}\siml \beta \siml 1$ (see previous section). This
range is mostly unconstrained by current observational restrictions on PBH mass
spectra \cite{carr05}. The choice of $t_1\sim 10^{-32}\s$ is then essentially
determined, if we want to explain both the current entropy  and the current 
dark matter.  
\begin{figure} [ht]
\centerline{\epsfxsize=3.4in \epsfbox{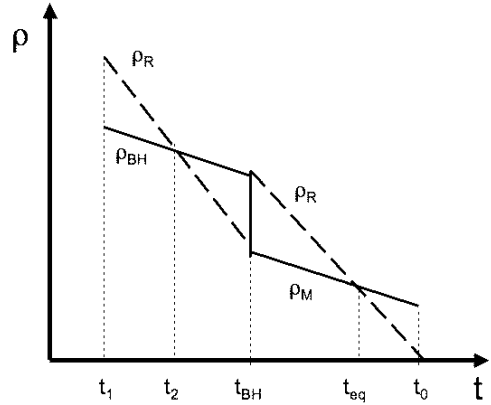}} \caption{The radiation, 
primordial black hole and subsequent dark matter evolution as a function 
of cosmological time. }
\label{fig:rho-t}
\end{figure}

\section{Baryon asymmetry}
\label{sec:barsym}

It is generally thought that the baryon asymmetry must have been  generated
by the epoch at which the Universe cooled below the electroweak energy scale
$T_{ew}\sim 300 \T300 \GeV$ \cite{tur79,bar91-92}. This corresponds,
extrapolating back from the present epoch, to a scale factor $a_{ew}\sim 2.6
\times 10^{-16} \T300^{-1}$ and an epoch $t_{ew}\sim 10^{-11.9}(\Omh)^{-2}
\T300^{-2}~\s$. This electroweak transition epoch $t_{ew}$ is, numerically, 
essentially the same as the evaporation epoch $\tbh\sim 10^{-11.4}\s$ defined 
in equation (\ref{eq:tbh}), for PBHs of mass $M_{BH}\sim 10^{6.6}~\g$ 
[eq. (\ref{eq:Mbh})] which formed at the epoch $t_1\simeq 10^{-32}~\s$.

The coincidence between the PBH evaporation timescale $\tbh$ and the 
electroweak timescale $t_{ew}$ is remarkable. The first is derived
from a requirement to explain the reheating and the observed entropy/DM
ratio starting from an inflationary early universe scenario, while the 
second is determined from a particle physics energy scale and the more recent
dynamics of the Universe. The fact that the Universe is baryon asymmetric
provides, in fact, an additional constraint on the epoch $\tbh$ at which PBHs 
evaporate, if the baryon asymmetry arises from baryonic decay products of PBH 
evaporation, which is manifestly CP-violating process.

The requirement that any net baryon number $n_B=n_b-n_{\bar b}$ produced in 
PBH evaporation should not be washed out by CP-violating currents expected 
at the electroweak transition epoch $t_{ew}$ imposes the requirement $\tbh 
\simg t_{ew}$.  This additional requirement is in fact satisfied by the 
entropy to dark matter constraint (\ref{eq:massratioteq}), requiring 
$\tbh\sim 10^{-11.4}\s$, which in turn constrains, through equation 
(\ref{eq:Mbh},\ref{eq:tbh}), the horizon entrance time of the PBH perturbations 
to be $t_1\simeq 10^{32}\s$. In addition, in order for the energy density of 
these PBHs not to be diluted to a negligible value, they must be born after 
the end of inflation, $t_1 \simg t_{end}$. The double constraint on $t_1 
\simeq 10^{-32}\s$ from the reheating/entropy/DM requirement on the one hand, 
and on $t_2 \simg t_{ew}$ to preserve any baryon asymmetry on the other hand, 
in turn constrains the end of inflation to occur at $t_{end}\simeq t_1\simeq
10^{-32}\s$.  In fact, either one of the previous two constraints acting 
individually (as long as PBHs are responsible either for the entropy/dark 
matter ratio or the baryon asymmetry) is enough to require $t_{end}\simeq 
10^{-32}\s$. The fact that both independent constraints acting simultaneously 
require the same value of $t_{end}$ is again remarkable.

The baryogenesis mechanism of \cite{bar91a} included mechanisms occurring 
at GUT temperatures, assuming that PBHs with $T_{BH} > 10^{14} \GeV$ radiate 
bosons which decay into a net baryon number. (A different PBH baryogenesis
mechanism in ekpyrotic/cyclic models \cite{Khoury:2001wf} has been discussed
by \cite{baumann04}).
However, any GUT scale baryon asymmetry can be washed out by Sphaleron 
processes during during the electroweak phase transition.  Sphalerons are 
non-trivial topological field configurations which generate a net $B-L$ 
number. It was shown by Cohen, Kaplan and Nelson \cite{cohen} that it is 
possible to use the electroweak sphaleron (instantons) to generate baryon 
asymmetry through the well known ABJ anomaly equation:  
\beq 
\partial_{\mu}J^{\mu}_{B}= N_{f}(\frac{g^{2}}{32\pi^{2}}W\tilde{W}-  
 \frac{g'^{2}}{32\pi^{2}}B\tilde{B}) 
\label{eq:anomaly} 
\enq 
where $N_{f}$ is the number of families, $W_{\mu\nu}$ is the weak field
strenght, $B_{\mu\nu}$ is the hypercharge field strength and $g$ and $g'$ 
are the gauge couplings.  The CKN mechanism states that baryogenesis can 
be spontaneous in the sense that a derivate coupling between a scalar field 
and the baryon number current is induced in general:
\beq 
\cal{L}\rm_{ind} = \partial_{\mu}\phi
N_{f}[\frac{g^{2}}{32\pi^{2}}Y_{CS}(A_{SU(2)}) +
  \frac{g'^{2}}{32\pi^{2}}Y_{CS}(A_{U(1)_{Y}})] 
\label{eq:lagrangian}
\enq 
where in general $F(A)\wedge F(A) = dY_{CS}(A)$, and substituting from eq 
(\ref{eq:anomaly}) we get 
\beq 
\cal{L}\rm_{ind} \simeq \partial_{\mu}\phi J^{\mu} 
\enq  
However the coincidence of PBH formation during and before the electroweak phase
transition temperature gives us a clue as to the origin of this field $\phi$. 
If the field $\phi$ is associated with the phase of the Higgs field we may be 
able to naturally generate the baryon asymmetry.  Indeed such a mechanism was 
made concrete by Nagatani \cite{nagatani99}.  In this mechanism the Higgs field 
forms a spherical domain wall around the PBH due to spontaneous electroweak 
symmetry breaking.  The gradient in the domain wall is the CP violating phase 
which also acts as the chemical potential to generate the net baryon asymmetry 
due to sphaleron processes.  The domain wall configuration is expressed as:
\begin{eqnarray}
 \langle\phi_1^0(r)\rangle &=&
  \left\{
   \begin{array}{lcl}
    0 & & (r \leq r_{\rm DW}) \\
    v_1 f(r) \; e^{-i \Delta\theta (1-f(r))} & & (r > r_{\rm DW})
   \end{array}
  \right.,
\end{eqnarray}
where $f(r) = \sqrt{1 - (T(r)/T_{\rm weak})^2}$, $T(r)$ is the local 
temperature measured at a radius $r$ from the black hole center. This
temperature 
gradient is determined by the radiation energy density gradient produced by the 
radiation outflow from the black hole.

In this configuration of the Higgs vacuum expectation value, the width of the
domain wall $d_{\rm DW}$ is equal to the depth of the symmetric region.  The 
Hawking radiation (particles) emanating from the black hole traverse this domain
wall, and the energy gradient in the wall induces a shpaleron transition which 
creates a net baryon number from the Hawking radiation\footnote{PBHs can also solve the domain wall and monopole problems according
to \cite{Stojkovic:2005zh,stokovic}.} .  The net baryon number 
$n_B$ resulting from this process can be calculated directly\cite{nagatani99}, 
and for PBH temperatures $\Tbh\sim 10^{6.5}-10^{7.5}\GeV$ the resulting net 
baryon  number, and the ratio of the net baryon to entropy (where the latter is
as calculated in the previous section) is $n_B/s \simeq 10^{-10}$, satisfying
the BBN constraints. %and WMAP results\cite{sper06}.
Remarkably, as shown in the previous two sections, this temperature is
essentially the  same temperature (\ref{eq:Tbh}) corresponding to PBHs formed at $t_1\sim 
10^{-32}~\s$ and evaporating at $\tbh\sim 4\times 10^{-12}~ \s \sim t_{ew}$, 
which can produce both the observed entropy and the observed dark matter
density.  These same PBHs can therefore also produce the right net baryon 
number and the entropy per baryon of the universe.

\section{Discussion}
\label{sec:discussion}

The possibility that three major observational parameters of the universe,
namely the entropy density, the dark matter density and the net baryon to 
entropy ratio, may be simultaneously explained by a single mechanism is 
remarkable. In this scenario the reheating and the entropy is produced by 
the evaporation of promordial black holes.  If, as has been widely 
surmised, these leave relics whose mass is of order the Planck mass per 
evaporating black hole, these can provide the dark matter density.
For one or both of the above to come out right, the PBH mass must be in 
the ton range ($1\times10^{3} kg$).  A newer element, in addition to the above, is that 
the difference between the PBH temperature and the temperature of the 
universe provides a temperature gradient, through which evaporating 
particles can undergo CP-violating transitions leading to a net baryon 
number. A specific PBH evaporation domain wall mechanism can give the
observed net baryon number and baryon to entropy ratio, in agreement 
with BBN constraints, when the PBH temperature is in the PeV range, 
corresponding again to the ton mass range.

The entropy, by itself, could in principle be produced by a range of PBH 
collapse times $t_1\siml 10^{-32}\s$. However, the additional requirement 
of relating also the dark matter density, the net baryon number, or both, 
to the evaporation process narrows the PBH formation time to the epoch 
$t_1\sim 10^{-32}\s$. Since PBHs with a significant  energy density must
arise at or after this epoch, this can be identified with the end of
inflation. This same epoch occurrs independently in hyperextended models 
of inflation where the non-minimal coupling is something other than 
quadratic in the scalar  field, leading also to a prediction 
\cite{Kosowsky:1992rz} of a gravitational wave background.

The PBH formation epoch $t_1\sim t_{end}\sim 10^{-32}\s$ also determines 
the reheating temperature $T\sim 300\GeV$, caused by the PBH evaporation 
at the epoch $\tbh\sim 10^{-12}\s\sim t_{ew}$.  The triple coincidence 
discussed here, based on the physics of PBH evaporation, provides a strong 
incentive for identifying PBHs as responsible for three of the key 
parameters of cosmological models,  namely the current entropy, the dark 
matter, and the net baryon asymmetry with the right baryon to entropy 
ratio. It also provides an upper limit for the end of inflation at the 
epoch $t_{end}\sim 10^{-32}\s$.  

{{\it Acknowledgements:} We are grateful to M. Peskin, M.J. Rees and 
P. Steinhardt for valuable comments, and to NSF AST0307376 for partial support.}

\end{document}